\documentclass[epj]{svjour}
\usepackage{graphics}
\usepackage{amsfonts}
\usepackage{amssymb}
\usepackage{graphicx}
\usepackage{epsfig}
\usepackage{color}
\begin{document}
\title{Bridging stylized facts in finance and data non-stationarities}
\titlerunning{Bridging stylized facts in finance and data non-stationarities}
\authorrunning{Camargo \and Duarte~Queir\'{o}s \and Anteneodo}
\author{Sabrina Camargo\inst{1} \and
S\'{\i}lvio M. Duarte~Queir\'{o}s\inst{2}\thanks{Corresponding author. sdqueiro@gmail.com} and
Celia Anteneodo\inst{3}
}
\institute{
Department of Physics, PUC-Rio,
Rua Marqu\^es de S\~ao Vicente 225, G\'avea, CEP 22453-900 RJ, Rio de
Janeiro, Brazil \vspace{.15cm}
\and
Istituto dei Sistemi Complessi --- CNR,
Via dei Taurini 19, 00185 Roma, Italy \vspace{.15cm}
\and
Department of Physics, PUC-Rio and
National Institute of Science and Technology for Complex Systems,
Rua Marqu\^es de S\~ao Vicente 225, G\'avea, CEP 22453-900 RJ, Rio de
Janeiro, Brazil
}
\date{Received: date / Revised version: date}
%
\abstract{
Employing a recent technique which allows the representation of nonstationary
data by means of a juxtaposition of locally stationary paths of different
length, we introduce a comprehensive analysis of the key observables in a
financial market: the trading volume and the price fluctuations.
From the segmentation procedure we are able to introduce a quantitative description of statistical features of these two quantities,
which are often named stylized facts, namely the tails of the distribution of
trading volume and price fluctuations and a dynamics compatible with the U-shaped profile of the volume
in a trading section and the slow decay of the autocorrelation function.
The segmentation of the trading volume series provides evidence of slow
evolution of the fluctuating parameters of each patch, pointing to the mixing
scenario.
Assuming that long-term features are the outcome of a statistical mixture of
simple local forms, we test and compare different probability density
functions to provide the long-term distribution of the trading
volume, concluding that the log-normal gives the best agreement with the empirical
distribution.
Moreover, the segmentation of the magnitude price fluctuations are quite
different from the results for the trading volume, indicating that changes in the
statistics of price fluctuations occur at a faster scale than in the case of
trading volume.
} 
\maketitle
\section{Introduction}

In the last decades, the description of dynamic and statistical quantities
related to Finance has turned into an appealing subject for
the exploration of physical concepts beyond the scope they were originally
introduced to \cite{books}.
Much of the effort has been put upon shedding light on trust-
worthy mechanisms leading to the emergence of power-law distributions,
e.g., for the log-price fluctuations the distribution of which exhibits
an asymptotic scale-invariant form with slow convergence to the Gaussian,
with the Berry-Ess\'een theorem defining the upper limit of difference
between obtained and expected cummulative distribution functions~\cite{be}.
It is well established that the changes in the share price are triggered
by a myriad of factors (previous price fluctuations,
deviations from the target price, news, etc.) that make some people willing
to buy and some other to sell.
Moreover, activity of a financial market  is non-stationary~\cite{lux,bertram,scalas,marsili,livan}.
To
this trait it was assigned the origin of fat tails in financial observables
like the trading volume~\cite{tailvol,epl-vol,celia-vol},
which on its turn would imply fat tails in the price fluctuations and on
volatility as well~\cite{volvol}.
This scenario abides by the Wall Street heuristic law disseminated by Karpoff's work
that it takes volume to make price move~\cite{karpoff}.

In both Physics and Finance, the treatment of non-stationary data is often tackled assuming sets of
coupled stochastic differential equations representing different scales of evolution of the system,
which frequently pave the way to demanding solutions~\cite{multiscale}.
Mixtures of stochastic processes, e.g., compound
Poisson processes,  can also be considered \cite{scalas}.
However, allowing for the fact that
to fit real data some of these equations must have large relaxation times, the modeling of
non-stationary quantities can be efficiently simplified by considering that the
system is in a generic steady state regime and the data are well
described by a juxtaposition of intervals of length $\ell $ characterized
by few $N$ parameters $\left\{ \pi \right\} $~\cite{beck}. At the scale $\ell $, the parameters
are assumed constant, but in the long-term follow a certain probability
density function.
Within this approach, the length of the local patches $\ell $ is systematically
constant as well. This time independence can only be understood as the
first order of the juxtaposition approach, because it is unlikely that complex systems
are so well behaved in this respect. Furthermore, it is intuitive to think that a dynamics
for the length of the regions of local stationarity contains valuable information with respect to
overall features of the observable, \emph{e.g.}, the evolution of the correlations.
In addition, we can look at the outcome of the single scale proposal as a (new) mixture of `cut and pasted'
elementary scales that screens the actual statistical nature of the local
parameters, in the context of what was called superstatistics by Beck and Cohen \cite{beck-cohen}.

Recently, a work of ours introduced a non-parametric segmentation procedure,
dubbed Kolmogorov-Smirnov segmentation (KSS)~\cite{kss}, aimed at defining quasi-stationary
segments of varying length in non-stationary time series (see \ref{app_segmentation}).
Although the non-uniform segmentation of
non-stationary time series was not a brand new approach to the problem \cite{montoro}, KSS clearly outperforms
previous approaches to the problem that use either local moments testing or principal component analysis
in accuracy or fastness~\cite{compete1,compete2,compete3}.

Generically, the hypothesis that the length of the segments of local stationarity is not constant
sets the scene for a real dependence between the values of the local parameters $\left\{ \pi \right\} $
and the duration $\ell $ of the patches. Therefore, the long-term distribution of observable $\mathcal{O}$ within
this framework is given by
\begin{equation}
P\left( \mathcal{O}\right) = \int \ldots \int p\left( \mathcal{O};\left\{ \pi
\right\} ,\ell \right) \,p_{\pi }\left( \left\{ \pi \right\} ;\ell \right) \,
 p_{\ell }\left( \ell \right) \,d\pi _{1}\ldots d\pi _{n}\,d\ell
\label{gen-eq}
\end{equation}
where $p\left( \mathcal{O};\left\{ \pi \right\} ,\ell \right) $ represents
the conditional probability of having a value $\mathcal{O}$ given local
parameters $\left\{ \pi \right\} $ in a segment of length $\ell $. Assuming $%
p_{\ell }\left( \ell \right) =\delta \left( \ell -\lambda \right) $ we get the constant
$\ell $ case.

After obtaining clear-cut results on heart-rate variability and atmospheric turbulence~\cite{kss},
we investigate the impact of the non-stationary nature of financial time series in a large set
of \emph{stylized facts}.
Specifically, from a thorough characterization of the features of the trading volume at short time
scales (1 minute), we pitch at describing not only its statistical properties but also at introducing a
proper representation of price fluctuations from statistical properties of trading volume as first
endeavoured using daily data~\cite{osborne} and more recently essayed in~\cite{baltic} using coupled equations.
Concretely, our analysis focus on the dataset composed of price fluctuations, $r_i(t) \equiv S_i(t) - S_i (t-1)$,
and the trading volume, $V(t)$, of the 30 blue chip companies defining the Dow Jones
Industrial Average recorded at every 1 minute during the second semester of 2004. This corresponds to \emph{circa} $5\times 10^{4}$ data
points for each quantity of every stock $i$.  For the sake of handleness we have normalized the trading volume
of each stock by its average value over the span $v_i (t) \equiv V_i (t) / \overline{V _i} $. The price fluctuations (or returns) were
kept as defined.

\section{Heterogeneities in trading volume}

\subsection{Statistics of the patches}

As we want to describe established key facts and statistical features of financial markets
from trading volume, we start our analysis by applying the KSS
algorithm to this last quantity. Despite working nicely without the need for any additional constraint,
\emph{e.g.}, a lower bound for the size of the segments, we curbed the length $\ell $ to a minimum
of $30$ minutes. This is the time scale describing a first regime of
the autocorrelation function of the trading volume that was found not only for this
same data but also for data from other markets~\cite{celia-vol,epjb-vol}. It is worth mentioning that the
introduction of this lower bound does not affect the results we present hereinafter,
namely the typical length of the segments of quasi-stationarity.~\footnote{It just affects the distribution
for small $\ell $ but the asymptotic behavior is the same.}

\begin{figure}[tbh]
\begin{center}
\includegraphics[width=0.75\columnwidth,angle=0]{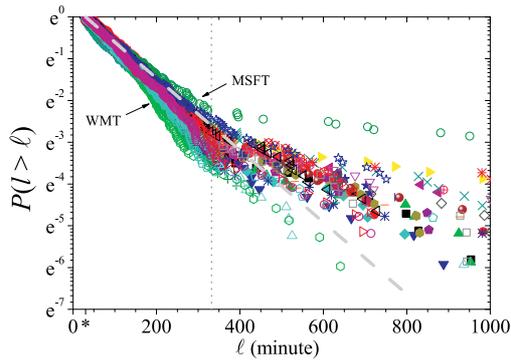}
\end{center}
\caption{ Complementary cumulative distribution function  $P\left( \ell \right) $\ vs segment
length $\ell $\ for the companies of the DJIA index in a $\ln $-linear
scale. Apart from the deviation in the tail a typical exponential decay is apprehended.
In the plot, we indicate the companies
with the shortest characteristic scale, Walmart (WMT), and the company with the longest characteristic scale,
Microsoft (MSFT).}
\label{fig-1}
\end{figure}

Let us first describe the probability density function (PDF) of the duration of the patches.
From a first visual inspection of Fig.~\ref{fig-1}, we noticed there is a well defined
exponential regime,
\begin{equation}
P_{\ell }\left( x > \ell \right) = \exp \left[ -\frac{\ell
-\ell _{\min }}{\lambda }\right] ,  \label{pdf-seg}
\end{equation}
which accounts for more that $95\% $ of the empirical distribution. Because
each firm presents as much as 300 segments, the remaining $5\% $ of the
empirical complementary cumulative distribution function  (EDF), which describes around 15 segments for each set, is strongly
affected by the finiteness of the patches set. It is thus tempting to consider the change in the
behavior of the curve a simple artifact. However, we do not have a random
modification. Instead, we observe a consistent decrease of its absolute value
for all the stocks and also that the changes come to pass at the same length $\ell \sim 330$ minutes.
Therefore, we reckon there exists a second regime in the length of stationary segments, which rules
the statistics of patches that last longer than a trading session.

Concentrating our efforts on the significant part of the distribution, we
used a log-likelihood adjustment procedure and from it we consistently found
that the EDFs fit for Eq.~(\ref{pdf-seg}) with similar values for all the stocks.
The the average value, $\left\langle \lambda \right\rangle $, is equal to $116\pm 12\,\,\min $, when
Microsoft (MSFT) is set aside ($\left\langle \ldots \right\rangle $ stands for
averages over companies).
For Microsoft, we have found a typical scale of 230 minutes, which is
quite different from the remaining values even when we compare it with
$\lambda =120$ minutes of Intel (INT), which is also traded at NASDAQ and that
agrees with $\left\langle \lambda \right\rangle $. The empirical distribution
function of $\ell $ is shown in Fig.~\ref{fig-1}. The reader should pay attention to the fact that despite
we did not remove overnight effects, which might affect a financial data analysis, our characteristic time scale
of local stationarity is significantly smaller than the span of a trading session. Moreover, should the trading
span influence our result, then there would be a separation between NYSE and NASDAQ traded stocks,
which is not the case bearing in mind Intel time scale.

The next logical step is to verify whether the length of the segments are related to one
another. This is appraised by looking into the behavior of the fluctuations,
\begin{equation}
\Delta \ell _{j}\left( i\right) =\ell _{i+j}-\ell _{i}.
\end{equation}
Already for immediate segments, $j=1$, we found white noise correlations,
\begin{equation}
C_{\Delta \ell _{1}}\sim \left\langle \Delta \ell _{1}\left( i+l\right)
\Delta \ell _{1}\left( i\right) \right\rangle -\left\langle \Delta \ell
_{1}\right\rangle ^{2}=\delta _{l,0} .
\end{equation}
However, when we analysed the correlation function of $| \Delta \ell _{1} |$,
we verified that it takes a lag around 4 segments to attain noise level, which using the value of
$\left\langle \lambda \right\rangle $ is close to the span of
a trading session.

\begin{figure}[tbh]
\begin{center}
\includegraphics[width=0.75\columnwidth,angle=0]{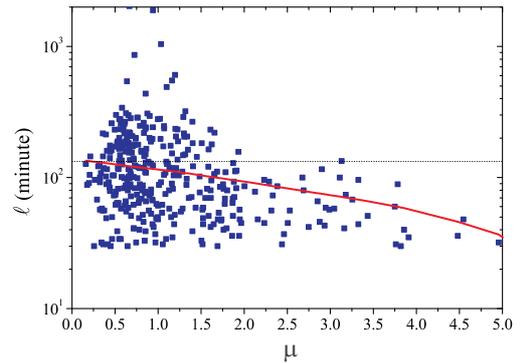}
\end{center}
\caption{ Typical dependence of the size of the segment of local
stationarity, $\ell $, vs local average value of the trading volume, $\mu$, for General Electric.
The line represents the local adjustment given by loess algorithm (see App. \ref{app_loess}). }
\label{fig-3}
\end{figure}

Considering high-frequency trading volume, it is known that markets tend
to exhibit high level of activity during the beginning and during the end of the
trading sessions~\cite{u-shape}. Therefore, the KSS \emph{must} yield indications of that U-shape
profile of the trading volume within a trading session, beyond the first indications
that the $C_{|\Delta \ell _{1} | } $ behavior is alluding to. We first looked
for a relation between the size of the segments, $\ell $, and its average
value of trading volume, $\mu_{\ell }$. Although the
plots $\ell $ versus $\mu $ are somewhat sprinkled (see Fig.~\ref{fig-3}), recurring to a standard
statistical technique of local regression (see App. \ref{app_loess}), we were able to
verify that there is an inverse relation $\ell $ and $\mu _\ell $
that goes beyond statistical error due to sample size. This result
is plausible because it is likely that periods with little activity (or small
$\mu $) last longer and that periods of high activity (or large $\mu $)
induce changes in the activity level more easily so that the local
stationarity condition is also more easily violated.

With the goal of  understanding how the segments length distributes within each
intra-day trading hour, we have looked at the
starting time of each segment of local stationarity and afterwards we coarse
grained them in such a way that the probability of obtaining a given band
is always equal to $1/8$. Accordingly, if the distribution of segments was
completely uniform along the day, then these probabilities would not vary
(within error bars).

\begin{figure}[tbh]
\begin{center}
\includegraphics[width=0.75\columnwidth,angle=0]{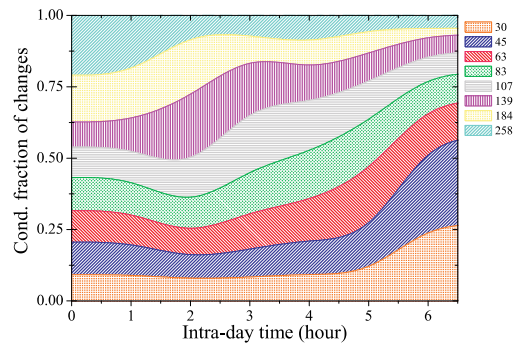}
\end{center}
\caption{Averaged stack column bar plot of the conditional probability of having a
change of local regime which lasts for $\ell $\ minutes averaged over all
companies and the frontiers are smoothed using a B-spline. The values of
the legend represent the initial value of each interval grouping.}
\label{fig-2}
\end{figure}

Taking into account the stack column bar plot in Fig.~\ref{fig-2},
we noted that there is in fact an intra-day dynamics for the conditional
fraction of the segments length. First, we apprehended that short and
long segments exhibit complementary behavior, \textit{i.e.}, longer segments
have their higher probability of starting during the first hours of a trading
session, perhaps reflecting trading sessions without much ado.
After that, it decreases to values smaller than $1/8$ from the second hour
of trading onwards, which indicates a strong intraday dynamics that moves on into
further sessions. For the smaller segments, we have almost the
same probability, which increases as the terminus of the session comes up. We relate this behavior
to the practice of cleaning the order book as the session approaches
its end. With a similar dependence there is a second group of segments of intermediate length,
but for which the final surging is very pronounced. The distinctive behavior with the trading time
can be already understood from these two analyses.

\begin{figure}[tbh]
\begin{center}
\includegraphics[width=0.75\columnwidth,angle=0]{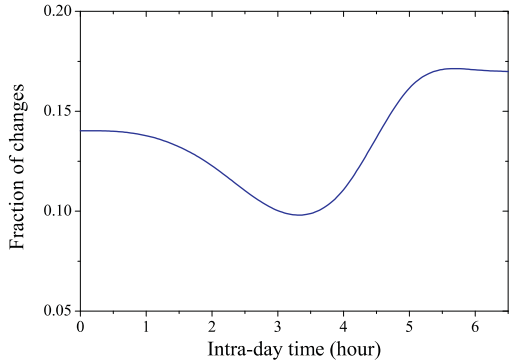}
\end{center}
\caption{Averaged probability of having a change of local stationarity for a given
intra-day time smoothed using a B-spline.}
\label{fig-4}
\end{figure}

Furthermore, to separate out the beginning of the session from the subsequent hours, we
appraised to what extent segments distribute within the trading session regardless
their length. Figure~\ref{fig-4} shows that changes of local stationarity
occur less in the middle of the session. Combining the analysis of
Figs.~\ref{fig-2}-\ref{fig-4} we perceive a dynamics totally compatible
with the aforestated U-shape in the trading volume.

\begin{figure}[tbh]
\begin{center}
\includegraphics[width=0.75\columnwidth,angle=0]{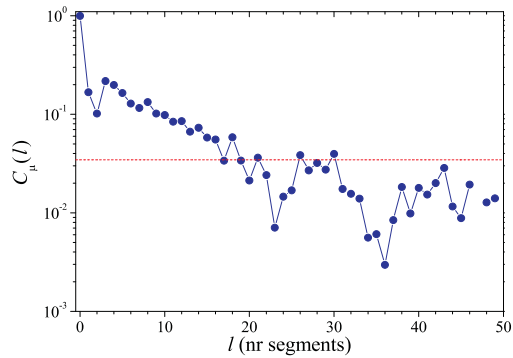}
\end{center}
\caption{ The averaged correlation function of the local mean value of the
trading volume vs the lag measured in segments. The horizontal dashed line
represents the noise level. Looking to the symbols we verify that $C_{%
\protect\mu }\left( l\right) $ reaches the noise level for a lag around five
days. Interestingly, we notice the intra-day signature for the bouncing back
of $C_{\protect\mu }$ at $l=3$ followed by other weaker and dwindling
rallies at $l=4$\ intervals until the noise level is attained. }
\label{fig-7}
\end{figure}

An important point in the mixing scenario is that of the slow evolution of
the fluctuating parameters that we fix within each patch. In Fig.~\ref{fig-7},
we show the average behavior of the correlation function of sequence of average
local values of the trading volume, $\mu $, values as a function of the lag
defined in number of segments units,
\begin{equation}
C_{\mu }\sim \left\langle \mu \left( i+l\right) \,\mu \left( i\right)
\right\rangle -\left\langle \mu \right\rangle ^{2}.
\end{equation}
Therein, it is visible that it takes as much as $18$ segments to
have the correlation at the noise level, which correctly accommodates
in the mixing approach.\footnote{We define the noise level as three
times the standard deviation of the correlation function when the
elements are shuffled.} Interestingly, we noted the existence of a
bounce back of the value of $C_{\mu }$ at $l=3$ which is dimly
repeated at $l=4$ intervals until noise level is reached. A similar curve is found when the
auto-correlation of the trading volume is analyzed. In other words, in segmenting the series
using the KSS algorithm, we preserved the long-term correlation function that also signals
the typical scale equal to the duration of a trading session which is close to 4 segments of average length.

\subsection{Long-term behavior from the local statistics of trading volume}
\label{statistics}

With the segmentation in hand and a first group of well-known properties
matching the segmentation results, we moved ahead into
probabilistic features. Owing to the assumption that the long-term
behavior is the outcome of a statistical mixture of simple local forms,
we assessed the statistical hypothesis that the trading volume is locally
described by one of these simple two-parameter PDFs: the $\Gamma $-distribution,
\begin{equation}
p_{\Gamma }\left( v;\left\{ \phi ,\theta \right\} \right) =\frac{v^{\phi -1}%
}{\theta ^{\phi }\,\Gamma \left[ \phi \right] }\exp \left[ -\frac{v}{\theta }%
\right] ,  \label{gamma}
\end{equation}
the log-Normal distribution,
\begin{equation}
p_{\mathrm{lN}}\left( v;\left\{ \phi ,\theta \right\} \right) =\frac{1}{%
\sqrt{2\,\pi }\theta \,v}\exp \left[ -\frac{\left( \ln v-\phi \right) ^{2}}{%
2\,\theta ^{2}}\right] ,  \label{lognormal}
\end{equation}
the inverse $\Gamma $-distribution,
\begin{equation}
p_{\mathrm{i}\Gamma }\left( v;\left\{ \phi ,\theta \right\} \right) =\frac{%
\theta ^{\phi }}{\,\Gamma \left[ \phi \right] }v^{-\phi -1}\exp \left[ -%
\frac{\theta }{v}\right] ,  \label{invgamma}
\end{equation}
and Weibull distribution,
\begin{equation}
p_{W}\left( v;\left\{ \phi ,\theta \right\} \right) = \frac{\phi }{\theta
^{\phi }}v^{\phi -1}\exp \left[ -\left( \frac{v}{\theta }\right) ^{\phi }%
\right] .  \label{weibull}
\end{equation}

We proceeded as follows: for every stock we have considered the segments
obtained by the KSS and looked for the best local fit for
PDFs~(\ref{gamma})-(\ref{weibull}) by means of optmizing the respective
log-likelihood function. Subsequently, we checked the
statistical significance of each fit considering the quantity
$\sqrt{\ell } \, d_{\max }$, where $d_{\max }$ is the maximum distance
between the EDF and the fitting cummulative distribution
function assuming a Lilliefors approach.\footnote{We opted by the
Lilliefors criterion instead of the standard Kolmogorov one in order to
check the difference between distributions on the left and on the right
of each value.}
From this procedure, we learnt that the log-normal distribution presents
the smallest value $\langle \sqrt{\ell } \, d_{\max } \rangle _{i} = 0.82 \pm 0.06$,
with the other distributions yielding average results greater than one
($\langle \ldots \rangle _{i}$ stands for average over all the
segments of company $i$).

Alternatively, having applied the Kolmogorov-Smirnov statistical distance
criterion with an $\alpha $-value equal to $0.05$ to each segment for
each testing distribution, we found an average ratio of statistical
significance equal to $0.95\pm 0.04$ for the log-Normal. For the remaining
test distribution we got a statistical significance ratio equal to
$0.78\pm 0.06$ for the $\Gamma -$distribution and $0.81\pm 0.04$ for the
Weibull distributions. Once more, the worst fit is for the inverse
$\Gamma $-distribution, which gave $0.41\pm 0.09$, \emph{i.e.}, a
performance ratio below $1/2$. The good results of the $\Gamma $-distribution
underpin the previous approach of a local Feller process~\cite{epl-vol,celia-vol,jstat-vol}.

An individual analysis of the companies also shows that the log-Normal tested as the best
local distribution for all the 30 stocks and the inverse $\Gamma $-distribution
the worst of the test hypotheses.

As previously denoted by Eq.~(\ref{gen-eq}), the long-term distribution is the result of a local
statistics, $p\left( v;\left\{ \phi ,\theta \right\} \right) $, that is
weighed taking into consideration the statistics of $\phi $ and $\theta $, $%
g\left( \phi ,\theta \right) $. Let us start reporting how $\theta $ is distributed.
To tackle this point, we compared the test distributions by computing $\sqrt{n}\, d_{\max }$, where $d_{\max }$
is again the maximal distance between the EDF and each of the complementary cumulative
distribution function after a log-likelihood adjustment and
$n$ the number of $\theta $ values in the set, \emph{i.e.},
the number of segments we obtained each stock. The averages over all the companies and
medians of $\sqrt{n}\, d_{\max }$ are,
\begin{equation}
\begin{array}{cccc}
& \left\langle \sqrt{n}d_{\max }\right\rangle & & (\sqrt{n}\widetilde{ d_{\max }}) \vspace{0.15cm}
\\
\Gamma \textrm{-distribution:} & 1.05\pm 0.58 & & 0.95 \vspace{0.05cm} \\
\textrm{inverse}\; \Gamma \textrm{-distribution:} & 1.39\pm 0.47 & & 1.35 \vspace{0.05cm} \\
\textrm{log-Normal:} & 4.94\pm 2.27 & & 5.09 \vspace{0.05cm} \\
\textrm{Weibull:} & 1.13\pm 0.54 & & 1.10 \vspace{0.05cm} \\
\textrm{inverse}\, \textrm{Weibull:} & 2.15\pm 0.54 & & 2.16%
\end{array}%
\nonumber
\end{equation}
showing that the  distribution which better describes the long
term behavior  is the $\Gamma $-distribution. Looking more attentively at the results,
we perceived different behavior for NYSE and NASDAQ traded stocks. For the former,
$p\left( \theta ;\left\{ \gamma ,\kappa \right\} \right) $
is best described by Eq.~(\ref{gamma}) with average values
$\gamma =32.8\pm 4.7$ and $\kappa =0.028\pm 0.004$ and medians
equal to $32.3$ and $0.028$, respectively.
Then again, for Intel and Microsoft, the best fit
$p\left( \theta ;\left\{ \gamma ,\kappa \right\} \right) $ is given by
Eq.~(\ref{weibull}) with similar exponent and scaling parameters for both the stocks, namely $\left\{
\gamma = 3.25, \kappa = 1.26\right\} $ and $\left\{ \gamma = 2.86, \kappa = 1.19\right\} $.
The values of the parameters $\gamma $ and $\kappa $ of NYSE
companies gave on average $\theta $ equal to $0.92\pm 0.14$ and a
standard deviation equal to $0.16\pm 0.02$ while for Intel and Microsoft we
have $1.13\pm 0.38$ and $1.06\pm 0.40$, respectively. It is worth remembering
that we have normalized our finite series of trading volume dividing each one by
its average value.

\begin{figure}[thb]
\begin{center}
\includegraphics[width=0.75\columnwidth,angle=0]{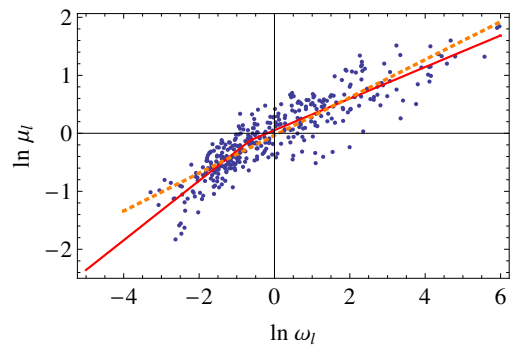}
\end{center}
\caption{ Scatter plot of $\ln $\ $\protect\mu _{l}$\ vs $\ln \,\protect%
\omega _{l}$\ for General Electric. The full line represents a numerical
adjustment with Eq.~(\protect\ref{crossovertheta}) and the dotted line is
a simple power-law.
}
\label{fig-5}
\end{figure}

The problem of the PDF of $\phi $ is very much simplified by
another empirical finding of ours. In performing a scatter plot of
the local average, $\mu _{l}\equiv \bar{v}_{l}$, versus the local variance,
$\omega _{l}\equiv \overline{v^{2}}_{l}-\bar{v}_{l}^{2}$, we perceived
a clear dependence between these two moments. Setting the
scatter plot in a $\ln -\ln $ scale, Fig.~\ref{fig-5} shows that this dependence is
close to a dual linear relation,

\begin{eqnarray}
\ln \mu _{l}  &=&\left\{ \alpha ^{\left( >\right)
}\ln \omega _{l} +\eta ^{\prime \left( >\right)
}\right\} \,\Theta \left[ \ln \omega _{l} -\Omega \right]   \label{crossovertheta} \\
& & + \left\{ \alpha ^{\left( <\right) }\ln \omega _{l} + \eta ^{\prime \left( <\right) }\right\}
\,\Theta \left[ \Omega -\ln \omega _{l} \right] ,  \nonumber
\end{eqnarray}
with,
\begin{equation}
\Omega \equiv \frac{\alpha ^{\left( <\right) }-\alpha ^{\left( >\right) }}{\beta
^{\left( >\right) }-\beta ^{\left( <\right) }},  \label{theta}
\end{equation}
where $\Theta$ is the Heaviside function and
$\eta ^{\prime \left( \gtrless \right) }\equiv \beta ^{\left( \gtrless
\right) }+\eta ^{\left( \gtrless \right) }$ (for the sake of conciseness we omitted
the dependence of $\mu _{l}$ and $\omega _{l}$ on $\phi $ and $\theta $). The variable
$\Omega $ represents the value at which we obtained a crossover and
$\eta ^{\left(\gtrless \right) }$ is Gaussian distributed with standard deviation
$\sigma _{\eta ^{\left( \gtrless \right) }}$ and null mean. For all the companies, except
3M, the crossover dependence Eq.~(\ref{crossovertheta})  was
found with  $\left\langle \mu _{l}\left( \Omega \right) \right\rangle =1.23\pm 0.88$.
This suggests the existence of a regime for smaller and another one for
larger trading volumes, as a previous scaling analyses suggested~\cite{scaling}.
Regarding the remaining parameters we obtained the following values
above and below $\Omega $,
\begin{equation}
\begin{array}{ccc}
\alpha: & 0.24\pm 0.08, & 0.45\pm 0.05; \vspace{0.05cm}\\
\beta:  & 0.22\pm 0.19, & 0.03\pm 0.11; \vspace{0.05cm}\\
\sigma _{\eta}: & 0.39\pm 0.08, & 0.28\pm 0.05 . \vspace{0.05cm}
\end{array}
\end{equation}
As regards the median,
which is less sensitive to extreme values we got:
$\widetilde{\alpha ^{\left( \gtrless \right) }}=\left\{ 0.23,0.44\right\} $,
$\widetilde{\beta ^{\left( \gtrless \right) }}=\left\{ 0.22,0.02\right\} $,
$\widetilde{\sigma _{\eta ^{\left( \gtrless \right) }}}=\left\{ 0.37,0.26\right\} $. Two notes on the
relation between $\omega _\ell $ and $\mu _\ell $ are still worthwhile: first, we tried adjusting
the scatter plot with a 2nd order polynomial, but the results were clearly
worse; second, although the dual
relation provides a better description of the data, a simple power-law adjustment fits the points fairly well, as shown
by the dotted line in Fig.~\ref{fig-5}.

For a log-Normal distribution defined by the parameters $\phi $ and $\theta $, the mean and the variance
are equal to,
\begin{equation}
\mu =\exp \left[ \phi +\frac{\theta ^{2}}{2}\right] ,  \label{meanlognormal}
\end{equation}
and,
\begin{equation}
\omega =\left( \exp \left[ \theta ^{2}\right] -1\right) \exp \left[ 2\,\phi
+\theta ^{2}\right] ,  \label{variancelognormal}
\end{equation}
respectively. Using these two equalities, we get for $\Omega =\pm \infty $,
\begin{equation}
\phi =\frac{\beta }{1-2\alpha }-\frac{\theta
^{2}}{2}+\frac{\alpha \,\ln \left[ \exp \left[ \theta ^{2}\right] -1\right]
}{1-2\alpha }.
\label{h-theta}
\end{equation}

While for the case of the dual-linear relation, it is not hard to obtain
the equation yielding the crossover parameters $\phi _{c}$ and $\theta _{c}$,
\begin{equation}
\left\{
\begin{array}{c}
\phi _{c}+\frac{\theta _{c}^{2}}{2}=\frac{\alpha ^{<}\beta ^{>}-\alpha
^{>}\beta ^{<}}{\beta ^{>}-\beta ^{<}} \,,\\
\\
2\,\phi _{c}+\theta _{c}^{2}\ln \left[ \exp \left[ \theta _{c}^{2}\right] -1%
\right] =\Omega \,,%
\end{array}%
\right. \vspace{0.1cm}
\end{equation}
it is  very hard  to obtain a simple expression analogue to Eq.~(\ref{h-theta}),
namely $\phi \left( \theta ;\phi _{c},\theta _{c}\right) $. Moreover, despite
the fact
that the fits using Eq.~(\ref{crossovertheta}) are better than a simple power-law and also that this dual approach also
helps verify previous results over disparities between small and large trading volume,
the approach based on Eq.~(\ref{crossovertheta}) drags in additional complications, particularly when one wants to apply
fast numerical integration methods such as the global adaptive strategy algorithm~\cite{numint}.

Bringing together all these findings, the long-term distribution of trading
volume is finally obtained performing the integration,
\begin{eqnarray}
P\left( v\right) &=&\int \int p\left( v|\phi ,\theta \right) \,f\left( \phi
,\theta ,\ell \right) \,d\phi \,d\theta \,d\ell  \\
&=&\int \int \int_{\ell _{\min }}^{\infty } F_{\ell }\left[ \mu \left( \phi ,\theta \right) \right] \,g\left( \theta \right) \,\,f_{\ell }\left( \ell \right) \, p\left( v|\phi ,\theta \right) \, f\left( \phi |\theta \right) \, d\phi \,d\theta \,d\ell \nonumber \label{mistura}
\end{eqnarray}
where,
\begin{equation}
p\left( v|\phi ,\theta \right) =\frac{1}{\sqrt{2\pi }\theta \,v}\exp \left[ -%
\frac{\left( \ln v-\phi \right) ^{2}}{2\,\theta ^{2}}\right] ,
\end{equation}%
expresses the local log-Normal dependence of the trade volume. The function
\begin{equation}
f\left( \phi |\theta \right) = \delta \left( \phi - \left[ \frac{\beta }{1-2\alpha }-\frac{\theta
^{2}}{2}+\frac{\alpha \,\ln \left(  \exp \left[ \theta ^{2}\right] -1\right)
}{1-2\alpha } \right] \right) \label{condprobphitheta}
\end{equation}%
 embodies the dependence between local average and local standard deviation.
The function $F_{\ell }\left[ \mu \left( \phi ,\theta \right) \right] $ is a
Dirac delta functional similar to Eq.~(\ref{condprobphitheta}) that allows
writing the length of a segment, $\ell $, as a function of local parameters $
\phi $ and $\theta $ via the local value of $\mu $ given by Eq.~(\ref
{meanlognormal}), namely
\begin{equation}
F_{\ell }\left[ \mu \left( \phi ,\theta \right) \right] \equiv \delta
\left[
\ell -h\left( \exp \left[ \phi +\frac{\theta ^{2}}{2}\right] \right)
\right],
\end{equation}
where the function $h\left( x\right) $ represents the fit for the loess curves
$\ell $ vs $\mu $ (see, e.g., Fig. 2) with its argument, $\mu $, substituted
for primary parameters $\phi $ and $\theta $ in accordance with Eq. (\ref{meanlognormal}).
According to what we said, the distribution of $\theta $ is given by,
\begin{equation}
g\left( \theta \right) =\frac{\theta ^{\gamma -1}}{\kappa ^{\gamma }\,\Gamma %
\left[ \gamma \right] }\exp \left[ -\frac{\theta }{\kappa }\right] ,
\end{equation}%
and finally $f_{\ell }\left( \ell \right) $ is given by Eq.~(\ref{pdf-seg}).

\begin{figure}[tbh]
\begin{center}
\includegraphics[width=0.75\columnwidth,angle=0]{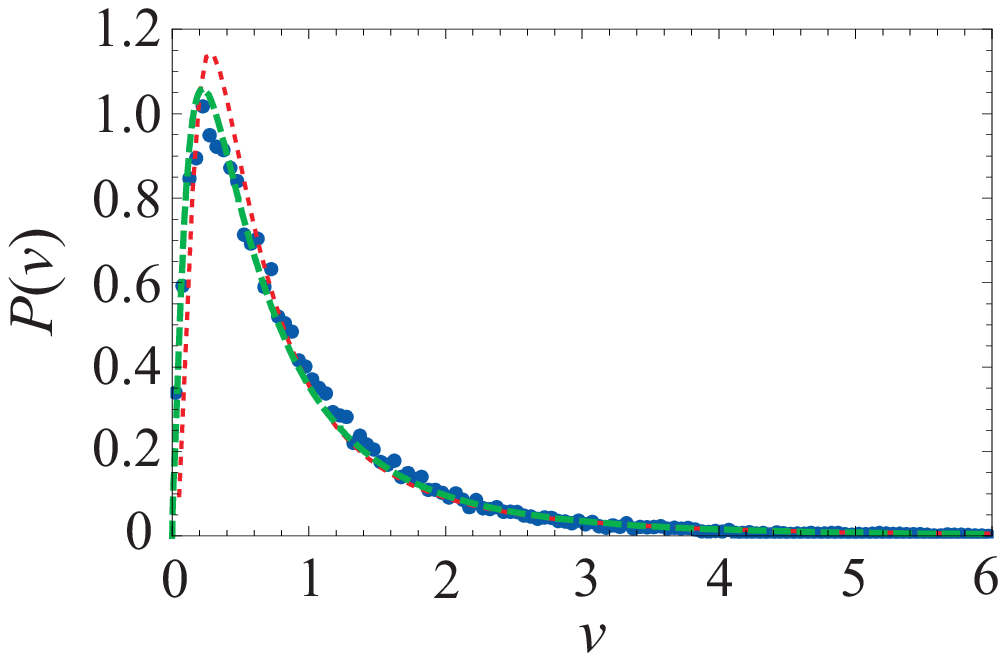}
\vspace{0.5cm}
\includegraphics[width=0.75\columnwidth,angle=0]{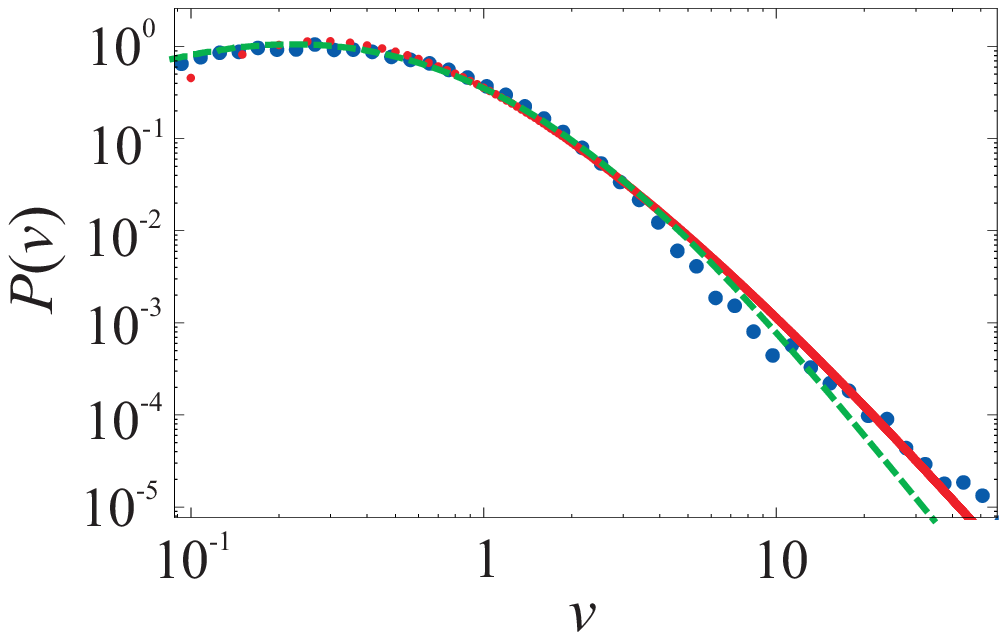}
\end{center}
\caption{ Long-term distribution function $P\left( v\right) $\ vs $v$. The
points the empirical PDF for the trading volumes of General Electric. The red dotted line
is obtained by numerically integrating Eq.~(\protect\ref{mistura}) assuming
the approximations described in the text and the green dashed line was
obtained using Eq.~(\protect\ref{mixingweighed}). The upper panel uses a $%
\log -\log $\ scale and the lower panel uses a linear-linear scale. }
\label{fig-6}
\end{figure}

Haplessly, the analytical determination of Eq.~(\ref{mistura}) is not possible
in this case. In respect of a numerical solution we can do it twofold.
In the first case, we assume that the length of a segment and the local moments are
independent and that the relation between $\ln \mu $ and $\ln \omega $ is linear.
Alternatively, one can appraise out the mixing scenario considering a weighed mixture
of the $n$ local log-Normal distributions defined by local parameters $\phi _i$ and
$\theta _i$, wherein the relative length of the $i$-th segment, $\ell _i / L$, plays the
role of the weight,
\begin{equation}
P\left( v\right) \simeq \frac{1}{L}\sum_{i=1}^{n}\ell _{i}\frac{1}{\sqrt{%
2\,\pi }\theta _{i}\,v}\exp \left[ -\frac{\left( \ln v-\phi _{i}\right) ^{2}%
}{2\,\theta _{i}^{2}}\right] ,  \label{mixingweighed}
\end{equation}
($L$ is the length of the time series). As understood in Fig.~\ref%
{fig-6}, despite the simplifications both approaches already yield a good agreement for small,
central and large values of the trading volume. This is particularly clear for the latter case in which we
basically do not assume any approximation. In this case, we also implicitly benefit of using information on the finiteness of
the data, whereas Eq.~(\ref{mistura}) assumes an infinitely long time series and neglects the local average -- segment length
relation, which explains the better results given by the green dashed curves in Fig.~\ref{fig-6}.

\subsection{Probing the relation between trading volume and price fluctuations at the mesoscopic scale}

\begin{figure}[tbh]
\begin{center}
\includegraphics[width=0.75\columnwidth,angle=0]{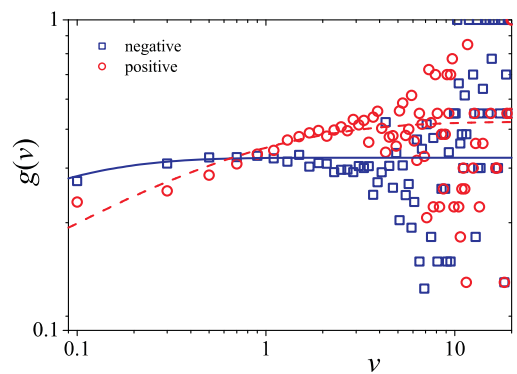}
\end{center}
\caption{Probability of having a positive (negative) price fluctuation vs trading volume. The points were
obtained for data of General Electric and the lines are the best fits using Eq.~(\ref{nao-nulo}).}
\label{fig-8}
\end{figure}

``It takes volume to make price move''~\cite{karpoff}. This assertion has been consistently
quoted and used as the starting point of attempts to establish a dynamical relation between trading volume and
price fluctuations~\cite{volvol,volvol2,scale}. While the famous adage is strongly supported and cultivated
by generations of brokers and econometricians, particularly those
who are interested in futures~\cite{futures}, the evolution to a stronger quantitative approach based on high-frequency
data analysis, especially the survey of order books, brought challenging explanations to the actual micro-mechanisms leading to the statistics of
price fluctuations~\cite{bfl}, namely the fact that large fluctuations are the outcome of differences between
ask and bid prices~\cite{farmer}. Nevertheless, the feeling that both plummets and significant increases are associated with large volumes remained,
in part due to the fact that historical slumps (at the daily scale) were accompanied
by large trading volumes and also because the cross-correlation function between price fluctuations and trading volume is above noise level.
Therefore, the natural question is: what is the real impact of trading volume on price fluctuations? Since fat tails in the
distribution of trading volume are mainly the
outcome of the heterogeneities in the activity (in our case in $\phi $ and $\theta $) we
can ask a slightly different question based on the classical works of Christie~\cite{christie} and Rogalsky~\cite{rogalski}:
to what extent are price fluctuations determined by non-stationarity of the trading volume?
From a probabilistic point of view, the simplest starting point to answer this question is to consider
Bayes' law,
\begin{equation}
\Pi \left( r\right) =\int p\left( r|v\right) P\left( v\right) dv.
\label{Preturnvolume}
\end{equation}
For highly liquid blue chip companies and 1 minute sampling rate, we definitely have
$P\left( v=0\right) = 0$. Nonetheless, it is possible that a given volume $v$ yields
no price fluctuation. To characterize the likelihood of this type of event,
we defined a probability,
\begin{equation}
g^{\left( 0 \right)}\left( v\right) \equiv 1-g^{\left( +\right)}\left( v\right) -g^{\left( -\right) }\left( v\right),
\end{equation}
where $g^{\left(\pm \right) }\left( v\right) $ corresponds to the probability of having a
positive (negative) price fluctuation for a trading volume $v$. The functions
$g^{\left( \pm \right) }\left( v\right) $ should verify two conditions: first, scraping
events like stock splits and dividends, when there is no trading volume the price
remains constant, \emph{i.e.}, $g (v^{(\pm )})(0) = 0$; second, for large values of $v$,
it most surely approaches a value independent of the trading volume, which is not necessarily
equal for negative and positive price fluctuations, as verifiable in Fig.~\ref{fig-8}.
For these reasons, we assumed that $g^{\left( \pm \right) }\left( v\right) $ is
fairly described by,
\begin{equation}
g^{\left( \pm \right) }\left( v\right) =G \tanh \left( \varpi \ v^{\beta
}\right) .
\label{nao-nulo}
\end{equation}
Averaging over all the companies we have,
\begin{equation}
\begin{array}{cccc}
& G & \varpi & \beta \vspace{0.15cm} \\
g^{\left( -\right) }: & 0.4\pm 0.03 & 2.56\pm 0.87 & 0.3\pm 0.1 \vspace{0.05cm} \\
g^{\left( +\right) }: & 0.47\pm 0.04 & 1.22\pm 0.29 & 0.25\pm 0.05. \vspace{0.05cm}%
\end{array}%
\nonumber
\end{equation}
These values, followed by a visual inspection of Fig.~\ref{fig-8}, point
that for small trading volume values the probability of having a negative value is higher than the
probability of a positive value with the relation between $g^{\left(
-\right) }$ and $g^{\left( -\right) }$ changing for $v \approx 1$. At first glance and taking
into consideration the risk-aversion ethos of financial agents, we would expect
precisely the opposite, \emph{i.e.}, small trading values dominating price rises and large
trading values associated with price decreases. However, minding the covariance
$\left\langle (r - \left\langle r \right\rangle)(v - \left\langle v \right\rangle) \right\rangle $,
we understand that this behavior corresponds to a high-frequency verification of
Ying's findings~\cite{ying} about the existence of a correlation between price fluctuations and
trading volume thus providing some quantitative support to the adage.

As regards trading volume associated with non-zero price fluctuations, it is
worth appraising the relation between volume and the price fluctuations,
\begin{equation}
\begin{array}{c}
\left\vert r_{t}\right\vert =\left\langle \ \left\vert r\right\vert \ |\
v_{t}\right\rangle +\eta _{t} \vspace{0.1cm} \\
=\mathcal{I}\left( v\right) +\eta ,%
\end{array}
\label{returnabsdef}
\end{equation}
where $\left\langle \, \left\vert r\right\vert \, |\ v\right\rangle $ represents
the expected magnitude of the price fluctuation produced by trading volume $v$.
We have represented this deterministic part of the relation between the
price fluctuations and volume by $\mathcal{I}\left( v\right) $ dubbing it
\textit{trading impact}. Although the term \textit{impact} has been introduced in
the context of order book analysis~\cite{impact}, it has been employed in
longer spells in which accumulated (meta) orders are considered~\cite{prx-impact}. At odds
with first proposals that assumed a linear relation between the
price difference and trading volume~\cite{kyle}, $S^{\prime }-S=\lambda ^{-1}\,v$ (with $\lambda $ being
the market depth), later approaches backed up by empirical analysis have
proposed that the long term $\mathcal{I}\left( v\right) $ is well described
by either power-law, $\left\vert r_{t}\right\vert \sim v_{t}^{\alpha }$, or
logarithmic, $\left\vert r_{t}\right\vert \sim \log v_{t}$, forms for the Paris and London
stock markets in the tick-by-tick~\cite{farmer,impactfunction} and 30 minute scales~\cite{hopman}.

In what follows, we tested the homogeneity of such
proposals,\textit{\ i.e.}, we aimed at finding whether the changes in the
local features of trading volume would impinge over its relation to the
price fluctuation. We carried out this approach twofold: we tried to
find a relation between the parameters describing the form of the trading impact
function and the size of the segments. Along these lines, we have tested
three different forms,%
\begin{equation}
\begin{array}{cc}
\mathrm{i)} & \mathcal{I}\left( v\right) =a+b\ln v_{t}, \vspace{0.05cm} \\
\mathrm{ii)} & \ln \,\mathcal{I}\left( v\right) =a+b\ln v_{t}, \vspace{0.05cm} \\
\mathrm{iii)} & \ln \,\mathcal{I}\left( v\right) =a+b\,v_{t}, \vspace{0.05cm} %
\end{array}
\label{impact}
\end{equation}%
where we have considered different versions of each one for positive and negative returns.
The power-law and logarithmic test functions are inspired by the previous work on impact functions and
the exponential, case iii) in Eq.~(\ref{impact}), is introduced because in complex systems it is ubiquitous
the emergence of power-laws from a mixture of exponential functionals with different
characteristic parameters. Since scatter plots of the price fluctuations with respect to
the trading volume are rather noisy at a local scale, we resorted once more to the loess
regression technique to describe cleaner curves.
Afterwards, we adjusted these curves using the expressions in Eq.~(\ref{impact})
and compared the $\chi ^{2}$ per degree of freedom values in order
to appraise which of the forms is the best. The average values for negative
and positive returns are the following,
\begin{equation}
\begin{array}{cccc}
& \chi ^{2}\,\,(\mathrm{negative}) & & \chi ^{2}\,\,(\mathrm{positive})
\vspace{0.15cm} \\
\mathrm{i)} & 4\times 10^{-4}\pm 0.002 & & 9\times 10^{-5}\pm 3\times 10^{-4}
\vspace{0.05cm} \\
\mathrm{ii)} & 0.006\pm 0.003 & & 0.006\pm 0.003 \vspace{0.05cm} \\
\mathrm{iii)} & 0.045\pm 0.033 & & 0.042\pm 0.017 \vspace{0.05cm} \,. %
\end{array}%
\nonumber
\end{equation}
Setting our sights on the best approach (smaller $\chi ^{2}$ values),
\textit{i.e.}, the logarithmic fit in Eq. (\ref{impact}), we have for the
negative returns $a=$ $0.056\pm 0.021$ and $b=0.0062\pm 0.004$ and for
positive returns $a=0.054\pm 0.015$ and $b=0.0059\pm 0.002$. In other words,
we have not found a significant difference between positive and
negative price fluctuations in respect of the trading impact. In addition,
further statistical analysis of $a$ and $b$ showed
that their distributions are significantly peaked.

\begin{figure}[tbh]
\begin{center}
\includegraphics[width=0.45\columnwidth,angle=0]{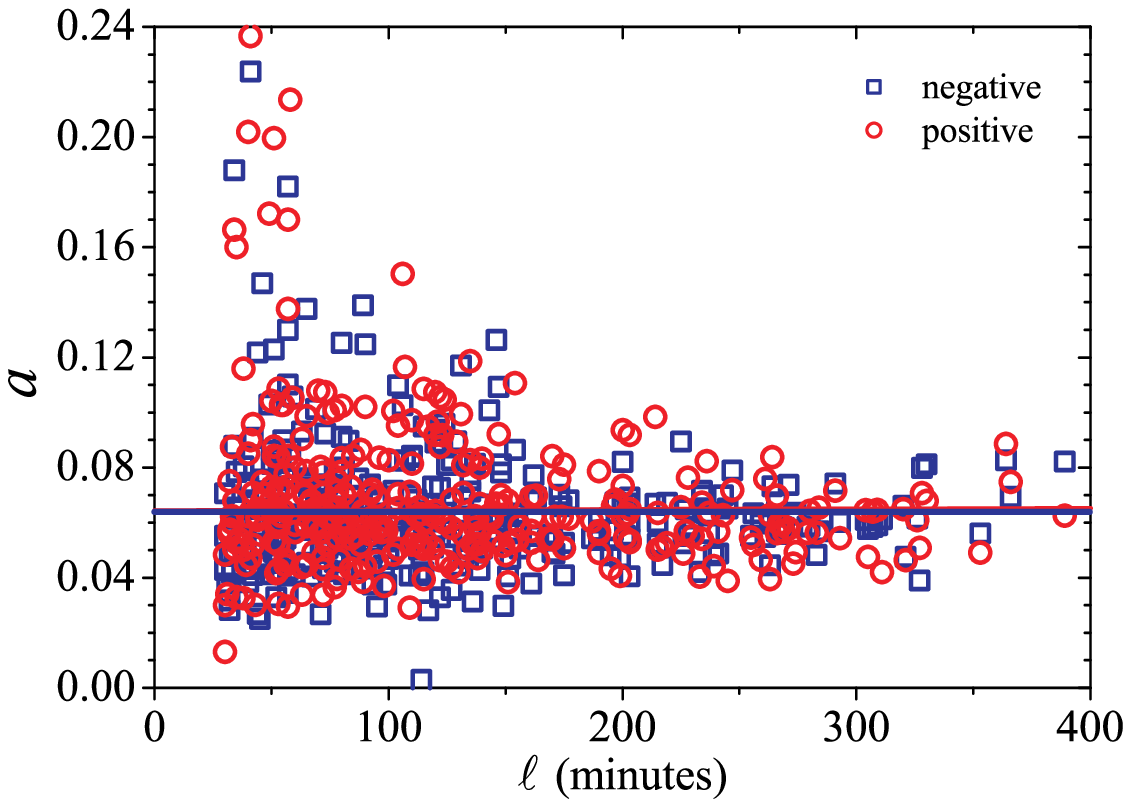}
\hspace{0.3cm}
\includegraphics[width=0.45\columnwidth,angle=0]{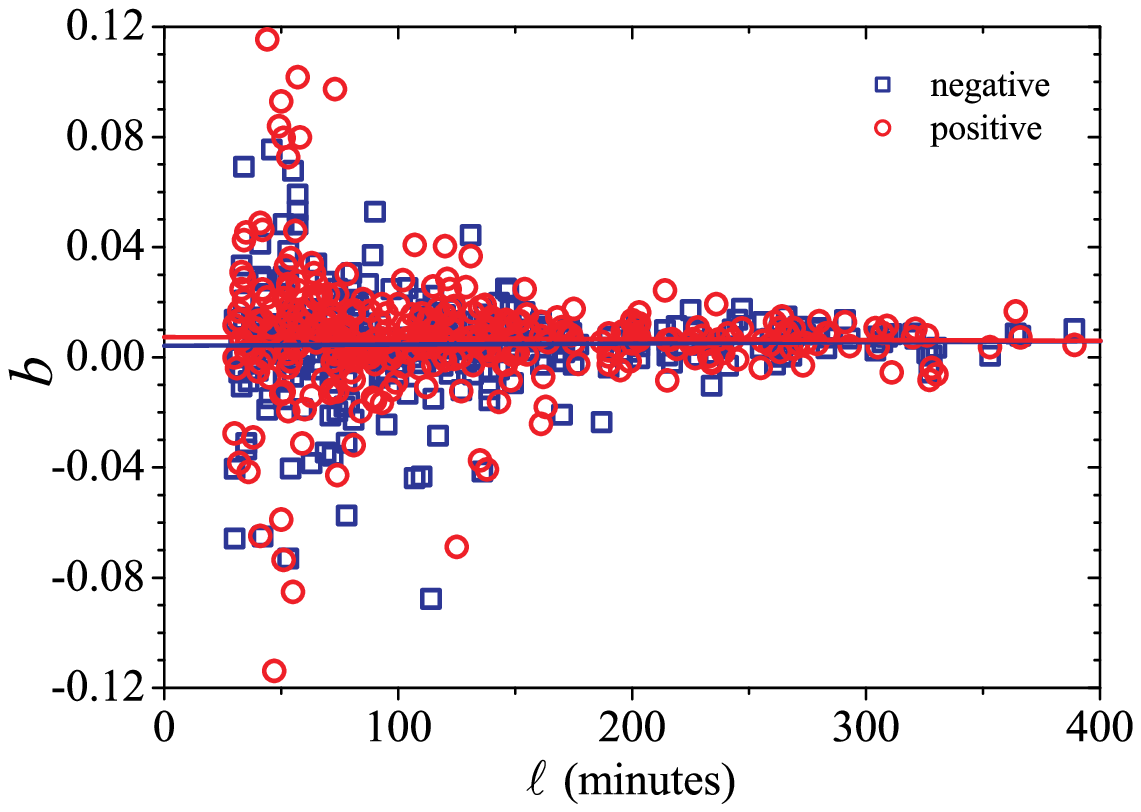}
\vspace{0.7cm}
\includegraphics[width=0.45\columnwidth,angle=0]{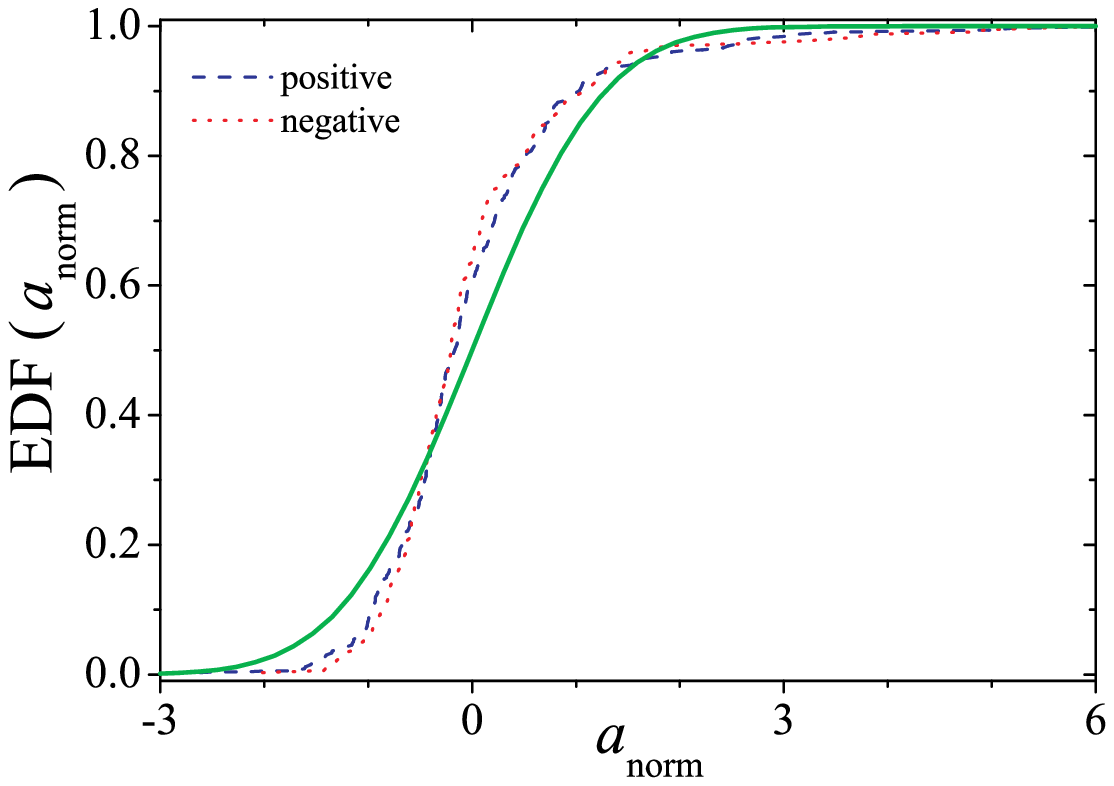}
\hspace{0.3cm}
\includegraphics[width=0.45\columnwidth,angle=0]{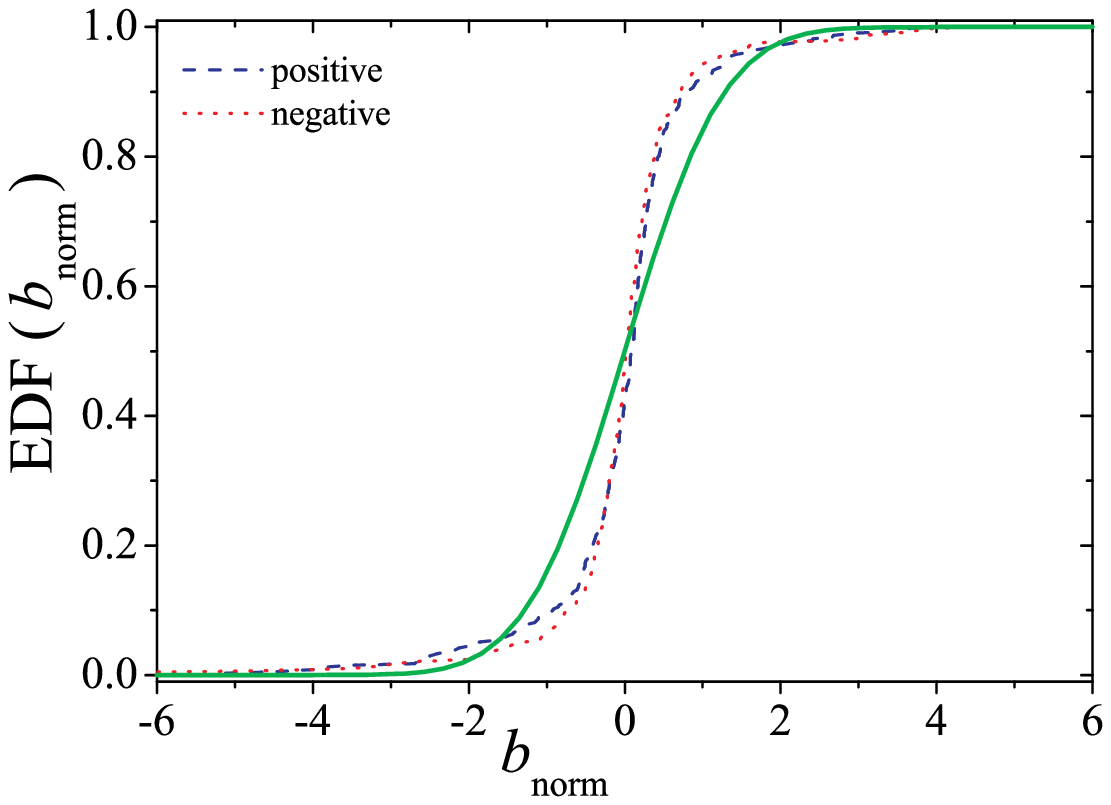}
\end{center}
\caption{Left: value of parameter $a$ in test i) of Eq.~(\ref{impact}) vs length of the segment $\ell $ (upper panel) and
EDF of detrended and normalized $a$ (lower panel) for each patch.
Right: the same but for parameter $b$. In the lower panels the green lines
represent the complementary cumulative distribution function of the Normal distribution showing that both $a$ and $b$ are not Gaussian distributed
in the long term. These data are for General Electric.}
\label{fig-9}
\end{figure}

Keeping our focus on the heterogeneities of the data, we further analyzed
whether there is a relation between the parameters $a$, $b$ and the size of
the segments, $\ell $ (see Fig.~\ref{fig-9}). Considering the linear adjustment of $a\left( \ell
\right) $ and $b\left( \ell \right) $ for positive and negative price
fluctuations, we verified they hardly vary with the segment length
yielding median slopes equal to $s_{a}=\left\{ -4.6\times 10^{-6},-4.3\times
10^{-6}\right\} $ and $s_{b}=\left\{ -7.7\times 10^{-7},2.8\times
10^{-6}\right\} $ with the same behavior verified using local regression.
Bearing in mind the magnitude of these slopes \textit{we assert that the trading impact functions
are homogeneous}. These two findings are represented in Fig.~\ref{fig-9}.

Let us finally introduce a simple argument which aims at explaining the
expected relation between return and volume. First, we describe the simple
case wherein a trading volume does not change the price. In this case, we
have in the long-term,
\begin{equation}
\Pi \left( r=0\right) =\int \left[ 1-g^{\left( 0 \right)} \left( v\right) \right] \ P\left(
v\right) \ dv,  \label{Pr0}
\end{equation}
where $P\left( v\right) $ is the long-term distribution Eq. (\ref{mistura})
[or Eq.~(\ref{mixingweighed}) in practical applications]. With respect to non-zero returns, the
conditional distribution has got a different form, namely,%
\begin{equation}
p\left( \,\left\vert r\right\vert \, |\ v\right) =f\left( \,\left\vert
r\right\vert \, |\ v\right) \, \left( g^{\left( +\right)}\left( v\right) + g^{\left( -\right) }\left( v\right) \right) ,  \label{Prneq0}
\end{equation}%
where $f\left( \,\left\vert r\right\vert \, |\ v\right) $ is the double conditional
probability of having a return of magnitude $\left\vert r\right\vert $ given
a trading volume $v$ that produces a non-zero price fluctuation.\footnote{%
In the last definitions we scrapped the distinction between positive and
negative returns for the sake of simplicity.} Allowing for Eq.~(\ref{returnabsdef})
and assuming that the error in the numerical adjustment, $\eta _{t}$, follows a
Gaussian distribution,
\begin{equation}
\mathcal{G} (\eta; \left\{ \left\langle \eta \right\rangle, \sigma _{\eta } \right\} ) =
\frac{1}{\sqrt{2 \, \pi} \, \sigma _{\eta }}\,
\exp \left[ - \frac{ \left( \eta - \left\langle \eta \right\rangle \right)^2 }{2 \, \sigma _{\eta } ^{2} } \right]
\end{equation}
we have,%
\footnote{%
Hereafter, we utilize the approximately equal signal because $f\left(
\left\vert r\right\vert \ |\ v\right) $ can only have a truncated form,
which is used to several problems, take into account that $\left\vert
r\right\vert >0$.}
\begin{equation}
f\left( \left\vert r\right\vert \, |\ v\right) \approx
\mathcal{G} \left(\left\vert r\right\vert ; \left\{ a+b\ln v , \sigma _{\eta } \right\} \right),
\end{equation}
and thus finally for $\left\vert r\right\vert \neq 0$ we get,
\begin{equation}
\Pi \left( \left\vert r\right\vert \right) =\int \mathcal{N}%
_{r}\left( a+b\ln v,\sigma _{\eta }\right) \, \left( g^{\left( +\right)}\left( v\right) + g^{\left( -\right) }\left( v\right) \right) \, P\left(
v\right) \, dv.
\label{pdf-vol}
\end{equation}
\begin{figure}[tbh]
\begin{center}
\includegraphics[width=0.75\columnwidth,angle=0]{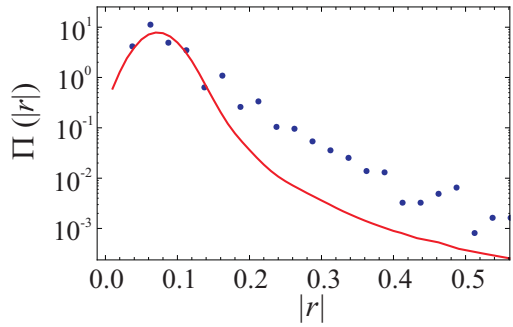}
\end{center}
\caption{ Probability of the magnitude of price fluctuations, $\Pi \left( \left\vert r\right\vert \right)$ ,
vs  magnitude of the price fluctuations, $ \left\vert r\right\vert $ for General Electric.
The red line was obtained by the (numerical) integration of Eq.~(\ref{pdf-vol}). The plot shows a factor 10
between data and the testing hypothesis.}
\label{fig-10}
\end{figure}

\begin{figure}[tbh]
\begin{center}
\includegraphics[width=0.75\columnwidth,angle=0]{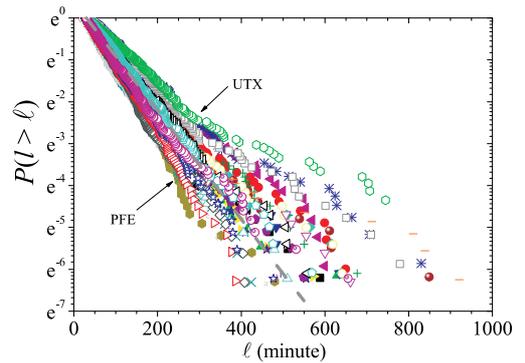}
\end{center}
\caption{Cummulative distribution of the segments of the segmentation of the
absolute values of the price fluctuations. Contrarily to Fig.~\ref{fig-1}, it is visible
a clear exponential decay and the average over characteristic times yields $77 \pm 15$ minutes.
The qualitative behavior among stock is also different. In this case the less non-stationary series
is United Technologies (UTX) whereas the most non-stationary is Pfizer (PFE).}
\label{fig-11}
\end{figure}

Plugging Eq.~(\ref{mixingweighed}) into Eq.~(\ref{pdf-vol}), we are
able to verify that although there is a relation between price fluctuations
and trading volume, it only yields a fair representation of the peak of the
distribution, which decays more slowly than the PDF from price -- volume arguments (see Fig.~\ref{fig-10}).
Taking into consideration that the peak
of a distribution concentrates the key part of the measure, we can state
that the heuristic adage relating trading volume and price fluctuations
is in some sense verified. \emph{However}, it completely fails at
describing the stylized fact concerning the slow decay of the distribution
$\Pi \left( \left\vert r\right\vert \right)$, as Fig.~\ref{fig-10} clearly shows.
So, what are the reasons for such misfire?
Within the context of the heterogeneous approach, we can understand the different
behavior of price fluctuations with respect to trading volume in applying the KSS
algorithm. Looking at the results we present in Fig.~\ref{fig-11},
we found that the length of the segments of local stationarity in
$|r|$ still follows a exponential like Eq.~(\ref{pdf-seg}), but with a much short typical
scale than the segmentation of trading volume, namely $\tau = 77 \pm 15$~minutes. This proves the different dynamics
of both quantities, particularly the respective degree of non-stationarity.
We still must remember that in the present approach, we wiped out factors like the fluctuations
of the parameters of the local impact functions that can be regarded as a proxy for the local volatility.
Accordingly, using a very different methodology our results go along the conclusion that large price
fluctuations are more about the volatility than the volume~\cite{scale,farmer-bouchaud}. We shall back to this point
in the Discussion.

\section{Mixing description of price fluctuations}

\begin{figure}[tbh]
\begin{center}
\includegraphics[width=0.75\columnwidth,angle=0]{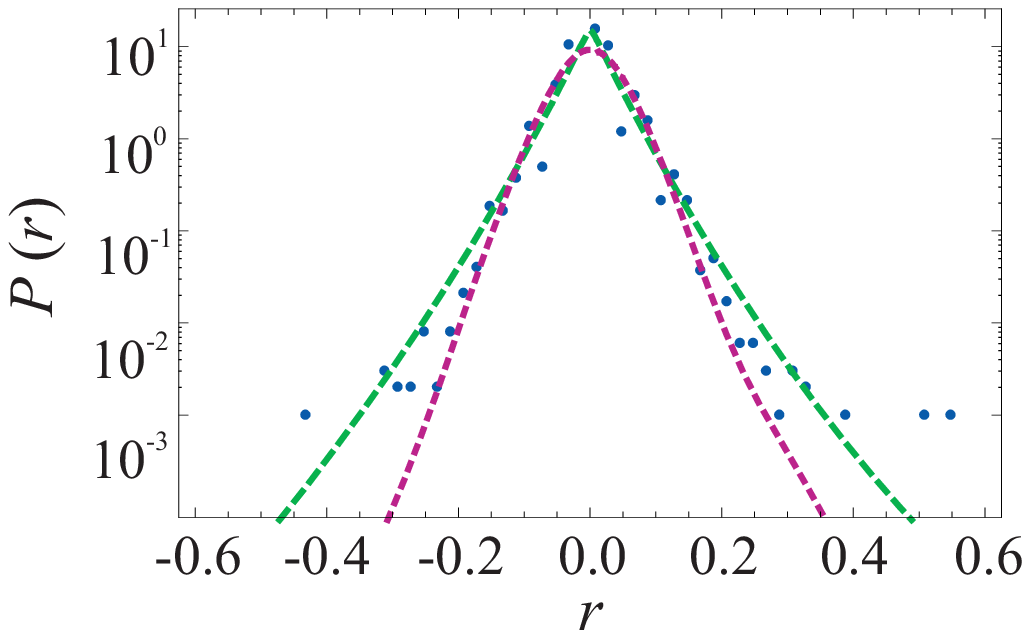}
\vspace{0.5cm}
\includegraphics[width=0.75\columnwidth,angle=0]{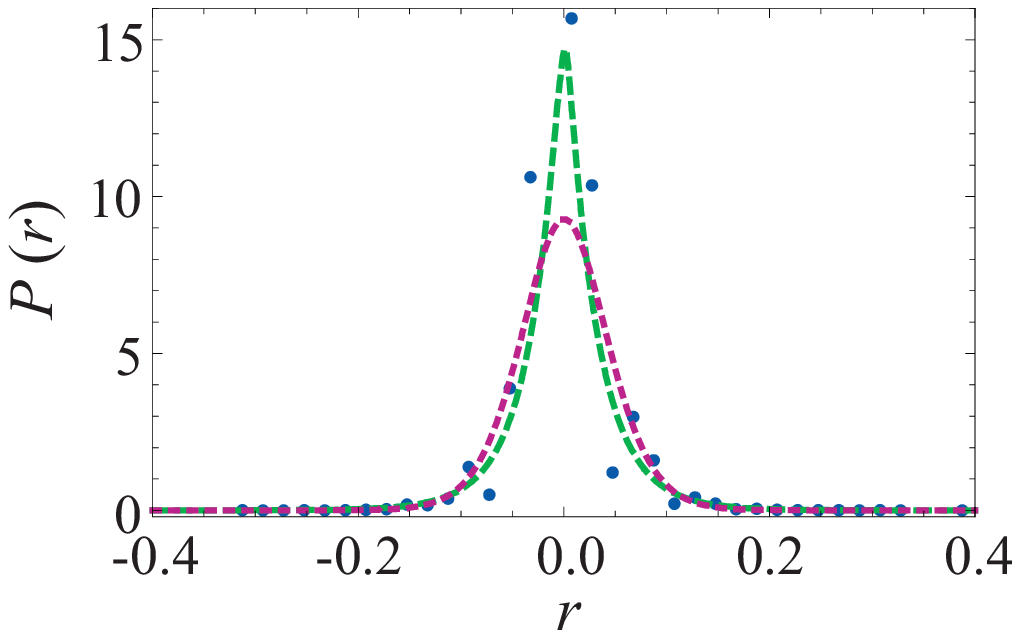}
\end{center}
\caption{Long-term distribution of the price fluctuations in log-linear and linear-linear scales (upper and lower panels, respectively).
The points correspond to the empirical PDF of General Electric. The green dashed line and the magenta dotted line
are the long-term distribution obtained from the segmentation procedure considering an exponential
and a Gaussian distribution. The parameters of the inverse-Gamma distribution of the squared volatility are
$\phi = 4.0$ and $\theta = 6.3 \times 10^{-3}$.}
\label{fig-12}
\end{figure}

Having verified that trading volume is not a relevant factor leading to fat tails
in the distribution of price fluctuations, we resort to the KSS in order
to milk some further information on the impact of the non-stationarity of the returns
in their local and long-term statistical properties. Traditionally, the
volatility has been justified by the impact of trading orders, but recent results at
the order book level as well as the results of our
previous section have shown that volatility actually reflects a raft of other things,
\emph{e.g.}, the random component in our trading impact among others. In our framework,
the first thing to do is to classify the statistical nature of the local standard deviation.
We assume as  \emph{local volatility} the variance of
the corresponding segment resulting from the segmentation
of the price fluctuations.
 Applying the same statistical
procedures of Sec.~\ref{statistics}, we verified that the best global distribution of the squared
volatility (local variance) is given by the inverse-Gamma distribution of Eq.~(\ref{invgamma}) with average
parameters $\phi = 2.5 \pm 0.7$ and $\theta = (4 \,  \pm 1)\, \times 10^{-3}$. For some companies the Gamma
distribution gave significant results as well.\footnote{These stocks are: American Express, Boeing, IBM, JP Morgan, Walmart.}
The observation of an inverse-Gamma distribution
for the squared volatility  is in accord with previous studies \cite{michiche} but it
 concurs with previous theoretical approaches aimed at justifying the
use of the Student-t (or $q$-Gaussian). However, this is just part of the story, to get such a
long-term distribution we still need to give statistical evidence that the price fluctuations are locally
Gaussian. Against the odds, we found that the local distribution is best locally described by an exponential
distribution,
\begin{equation}
p(r;{\mu ,\sigma }) = \frac{1}{2\,\sigma} \, \exp \left[ - \frac{ |r - \mu |}{\sigma } \right],
\end{equation}
for which the local average is equal to $\mu $ and the local variance is equal to
$\Sigma \equiv 2\, \sigma ^2$. Once again by employing the mixing indicated by Eq.~(\ref{gen-eq})
we obtain the long term distribution of the price fluctuations as presented in
Fig.~\ref{fig-12}.

In place of looking for full integration, we can simplify the calculation of $P(r)$ noticing that
the key deviation from the local exponential distribution comes from the large values of the volatility.
When $\Sigma \rightarrow \infty$ its distribution decays as $\Sigma ^{-1-\phi }$. Using the asymptotic
behavior in the integration rather than its full form we get $P(r) \sim |r|^{-1-2\, \phi}$.

The plots in different types of scales clearly show that the local Gaussian distribution does
not allow a good representation of the long-term distribution $P(r)$ both in the central
part and the tails. In addition, we verified that $P(r)$ decays almost as an exponential, which is compatible with
the large asymptotic exponent we obtained after using the values of $\phi $ found by fitting the local variance.

As a complement, we applied for each company the t-Student test \cite{numerical_recipes}
to verify whether the distributions of the local means of the returns were
compatible with a zero mean normal distribution. The $p$-values obtained
were $p<0.1$ for most companies; two companies with large $p$, namely Du Pont and McDonald's have
$p=0.39$ and $p=0.33$, respectively; and other five
companies (Caterpillar, IBM, Johnson \& Johnson, Altria and United Technologies) have $0.1<p<0.2$.
Concerning the skewness and the kurtosis of the distributions, the Jarque-Bera
\cite{jb} test showed a very good agreement with a normal distribution, providing for
all companies $p<0.001$.

\section{Discussion}
In this work we inspected the effectiveness of the statistical mixture approach
in mimicking primal financial quantities: price fluctuations and trading
volume. We proved that the proper segmentation of the time series, which considers
segments of varying length is fundamental for a correct description of
statistical and dynamical stylized facts. We did it employing a non-parametric
method of segmentation, the KSS~\cite{kss}, which assumes that differences
between segments can derive from discrepancies in any statistical moment, which define
the characteristic function of a probability density function.

Considering the trading volume, we reinforced the idea that a mixture of
distributions is able to nicely describe its long term PDF. Specifically, from our
results, the long-term PDF is effectually described by the statistical mixing
of juxtaposed stationary segments of unequal length wherein the trading volume
is log-Normal distributed. The distribution of the length of these segments is
dominated by a exponential form with a typical scale around $115$ minutes,
which decays much slowly for lengths greater than 330 minutes. This last regime mainly
represents the behavior of patches of stationarity that last
longer than a trading session. Bearing in mind stochastic mechanisms related to
the log-normal distribution, we can think about this functional form of the
trading volume as the result of a cascade of transactions,
$v(i) = \prod _{\tau = 1} ^{n_i} T _{\tau} $, with $\ln T $
representing the log of the size of the $\tau $-th trade that is Gaussian
distributed with average equal to $\mu $ and standard deviation equal t
$\theta $.

Locally, we also verified that there is an
intrinsic relation between the local average and the variance, \emph{i.e.}, between $\mu $
and $\theta $. With this relation, we were able to statistically
express the local behavior in terms of a single local Log-Normal parameter
[$\theta $ in Eq.~(\ref{lognormal})] that we learnt being well described by a
Gamma distribution. In addition, we noted that segments with large local
averages and small local averages behave differently and beyond
statistical effects, albeit the description considering a single behavior also
gives good results. These results are to be compared with the simpler approach of
segments of constant length. At the probabilistic level, the former case gives
a local distribution compatible with a Gamma distribution, which  provides a
fair local description of the data, with a $15 \% $ handicap though.
Notwithstanding, crucial differences arise in the description of the
data: first, there is an important relation between the local average and variance that would
not be learnt were we using fixed length segments or even applying a
segmentation method based on the analysis of the means; Second, and most
importantly, it would be impossible to capture and identify important dynamical
stylized facts such as the U-shape of trading volume within each session; the
slow decay of the autocorrelation function via the correlations of the
average value of trading volume in juxtaposed stationary patches as well as
the relation between the magnitude of the fluctuation of the segments length
which are significantly correlated within the trading session span.

Stemming from such a good description of trading volume, we were able to shed
light on the recent dispute between partisans of the famous relation between the
trading volume and price fluctuations popularized by Karpoff in~\cite{karpoff}
and new quantitative results that assign a minor role to volume in the dynamics
of price fluctuations. Our results indicate that each assertion has its own
domain of validity. On the one hand, we corroborated the claim conveyed
in~\cite{farmer,impact} that trading volume is not a key ingredient in large
price fluctuations. On the other hand, holding on the statistical properties
of trading volume we were capable of obtaining a fair representation of the
central part of the distribution of the magnitude of price fluctuations. This
finding agrees with the results of the cross-correlation between trading volume
and price fluctuations~\cite{volvol,volvol2,ying}, which are traditionally used
as the main argument to defend the intimate relation between price and volume.
However, our results clearly show that this cross-correlation (not greater than
$20 \% $) basically concentrates on small price fluctuations.
Within our framework, the most straightforward explanation for the short-coming
of the return --- volume relation is given by the segmentation of the magnitude
of the price fluctuations, the results of which are quite different from those
we obtained for the trading volume. Explicitly, for the magnitude of price
fluctuations we got a very clear exponential decay of the distribution of
segments of local stationarity without the slower tail exhibited by the distribution trading
volume segments. Quantitatively we found a typical scale around 75 minutes, which is
substantially smaller than the 115 minutes found in the segmentation of trading volume. Since
the length of the stationary segments acts as a simple, yet effective, way of
quantifying the extension of the non-stationary nature of a time series, we can
understand that changes in the impact of the volume or even in the probability
of having non-zero price fluctuations occur at a faster scale that is not
captured in the trading volume scale, leading to a faster decay of $\Pi (|r|)$.
The effects we have just mentioned can be combined and represented by the
volatility. Thence, working at a different scale our results prop up the
statement that ``there's more to volatility than
volume''~\cite{farmer-bouchaud}.
Complementarily, we might also say that the adage about the volume being
responsible for the price changes can be accepted in the same way Black-Scholes
equation is valid in option pricing: it gives a fair forecast during a good
part of the time, but it completely drops the clanger in the cases wherein
one can make (lose) big money.

Finally, we verified that the (squared) volatility of price fluctuations
evolves as an inverse-Gamma distribution, which perfectly tallies with the
mixture distribution hypothesis that assumes price fluctuations are locally
Gaussian distributed and which is the cornerstone of Engle's
ARCH model~\cite{engle}. Regardless, when we looked into the local distribution
of price fluctuations we concluded that they do not follow a Gaussian
distribution, but an exponential distribution instead. Were the local
distribution Gaussian, we would have had a long term empirical distribution
well described by a Student-\emph{t} ($q$-Gaussian), which is not the case for
both the central part and the tails. This result is interesting twofold:
\emph{a)} although the volatility process does not fit that of the Heston
model~\cite{heston}, the local exponential we found agrees with the short term
behavior of this model and \emph{b)} It prompts the study of different
definitions of noise in ARCH-like processes~\cite{arches}.

After bearing good fruit at the minute scale of stock trading, this method can
be put to use in other financial problems at order book scale and
enhance reasoning about other financial products. Concerning the former it
would be interesting to analyze which additional features could be captured in
the dynamical properties of individual agents previously studied by a comparison of the
local means~\cite{ibex}. We should underscore that in the case of
atmospheric turbulence~\cite{kss}, the test of the means is useless in the
evaluation of local stationarity. In respect of other financial products, we
can mention the dynamics of futures and other derivatives.

\begin{acknowledgement}
We would like to thank Olsen Data for having provided the data. SC and CA acknowledge CNPq (Brazilian agency) for partial
financial support. SMDQ  benefits from the financial support of the Marie Curie Intra-European Fellowships programme.
\end{acknowledgement}

\appendix
\section{KS-segmentation}
\label{app_segmentation}

Considering that a nonstationary time series can be split into stationary
segments, the aim of segmentation is to find the optimal positions to separate the
time series in such segments. In KSS, these positions are obtained by finding,
along the series, the maximal
\begin{equation}
D\equiv D_{KS}(1/n_{L}+1/n_{R})^{-1/2},
\label{eq:dks}
\end{equation}
where  $n_{L}$ ($n_{R}$) is the number of points to the left (right) of the hypothetical
cutting point and $D_{KS}$ is the Komogorov-Smirnov distance between the complementary cumulative
distributions of these two samples.
Once we find the  position  of maximal distance $D^{max}$,
we test the statistical significance (at a
chosen significance level $\alpha =1-P_{0}$) of a potentially relevant cut at that point.
That is achieved by comparison with the value of $D$ that would be
obtained was the sequence random. The critical value is given by the
phenomenological expression \cite{kss}
\begin{equation}
D^{max}_{crit}(n)=a(\ln n-b)^c,
\label{eq:dcrit}
\end{equation}
$(a, b, c)$ = ($1.41$, $1.74$, $0.15$), ($1.52$, $1.80$, $0.14$), and
($1.72$, $1.86$, $0.13$) for $P_0=0.90, 0.95, 0.99$, respectively.
If $D^{max}$ exceeds the critical value for the selected significance level $D^{max}_{crit}(n)$,
then the cut is done.
The procedure is then recursively applied starting
from the full series, until no segmentable patches are left. See~\cite{kss} for further details.

\section{Loess}
\label{app_loess}

In order to have a smooth set of points from scattered data $(x_i,y_i)$, $i=1,\dots, n$, we apply the
robust locally weighted regression (loess) \cite{loess} to obtain the estimated values for each
point.
The procedure consists of two parts. First, the weight
function depends on the distance to the $r$-th nearest neighbor
and a weighted least-squares fitting procedure  gives the estimated values
for each point. Second, a new factor is introduced in the weighting
computation, based on the residuals of the first fitting procedure,
improving the weights in the sense that large residuals will have small weights and
small residuals will have large ones.

We can summarize the loess procedure as follows: for each point $i$, we compute the distance
$h_i$ from $x_i$ to its $r$-th nearest neighbor. The $k=1,\dots, n$,
(with $k\not= i$) weights for each point $x_i$ will be given by
\begin{equation}
\omega_k(x_i)=W\left(\frac{x_k-x_i}{h_i}\right),
\label{eq:weight}
\end{equation}
where $W$ is the tricubic weight function
$$
W=\left\{
\begin{array}{cc}
(1-|x|^3)^3, & |x| < 1\\[0.5cm]
0,  & |x| \ge 1.\\
\end{array}
\right.
$$
Then, in our cases, a linear least-squares fitting with weights given by Eq.~(\ref{eq:weight})
determines the estimated $\hat{y}_i$ that corresponds to $x_i$ and its
residual, $e_i=y_i-\hat{y}_i$. A different set of weights,
$\delta_k=W(e_k/(6\,s))$,
is defined for each $(x_i, y_i)$ based on the size of $e_i$, and $s$ is the
median of $|e_i|$. The new estimated values are obtained as before but
with $\omega_k(x_i)$ replaced by $\delta_k\omega_k(x_i)$. This calculation of
$\delta_k$ is iterated as much as necessary to have a satisfactory smoothed
curve, for example when $s$ stabilizes.
Further discussions and examples are presented in~\cite{loess}.

\end{document}